\documentstyle[twocolumn,epsf,prl,aps]{revtex}
\newcommand {\be}{\begin{equation}}
\newcommand {\ee}{\end{equation}}
\newcommand {\bea}{\begin{eqnarray}}
\newcommand {\eea}{\end{eqnarray}}
\begin{document}
\draft

\title
{Conductivity of Layered Crystals}

\author{G. A. Levin and C. C. Almasan}
\address{Department of Physics, Kent State University, Kent  OH 44242\\
\vspace{.5 cm}
{\rm \begin{quote}
We show that the resistive anisotropy  of an anisotropic medium
is determined by the  ratio of the phase coherence lengths.
In layered crystals in which the interlayer
transport is incoherent,  the out-of-plane  phase coherence length is fixed
and temperature independent. This leads to a temperature dependent resistive
anisotropy  and to the coexistence of metallic in-plane and
non-metallic out-of-plane conductivities.  Our approach provides a
description of the c-axis conductivity in the highly nonclassical regime,
characteristic of layered cuprates.
\end{quote}
}
}
\maketitle


The normal state of highly anisotropic crystals such as
high-$T_c$ cuprates  exhibits a number
of unusual features. The coexistence of metallic in-plane ($\rho_a$) and
non-metallic
out-of-plane ($\rho_c$) resistivities within a range of
doping is one.\cite{Cooper}.
A crossover from metallic to non-metallic temperature (T) dependence
is another\cite{Ando,Ando1}. From a theoretical point of view,
there is a great deal of interest in exploring the consequences of the idea that
interlayer transport in these crystals is incoherent, so that successive
interlayer
transitions  are uncorrelated as evidenced by the analysis
of  dc and infrared  conductivities\cite{Cooper,Basov}.

The coherence of the wave function of the charge carriers
can be  described in terms of the  phase coherence length, hereafter called the
Thouless length (TL),
the distance electrons travel between dephasing {\it inelastic}
collisions\cite{Lee}.
An anisotropic medium, naturally,  is characterized by an
anisotropic TL. Below we  show that, in general, the anisotropy
$\sigma_a/\sigma_c$ is determined by the ratio of  TL in the respective directions.
The incoherence of  the out-of-plane transport in layered crystals
implies that  TL in the c-direction is temperature independent, equal to the
spacing between neighboring layers. As a result,  the temperature dependence of
$\rho_c/\rho_a$ is completely determined by the T-dependence of the in-plane TL.
This  can  lead  to coexistence of metallic $\rho_a$ and non-metallic
$\rho_c$.
We also explore the consequences of the idea
that the evolution of the  conductivities ($\sigma_a$ and
$\sigma_c$) with temperature and doping can be described through their
dependence on the in-plane Thouless length.
One of the most interesting conclusions
that can be drawn from this analysis is that
there may exist a unifying description of the conductivity of layered
crystals  at different doping levels.

Let us consider an anisotropic medium characterized by three different TL:
$\ell_x\;,\ell_y$, and $\ell_z$.  The phase-coherent volume  of such a
medium, which may be
considered as a block
with sizes equal  to the respective TL (hyperblock)  has 
{\it isotropic conductance}; i.e., $g_x=g_y=g_z\equiv g$.
One way to prove this is by using the original Thouless
idea\cite{Thouless,Abrahams}
that the conductance of a block is
\be
g_i= \frac{e^2}{\hbar}\frac{dN}{dE}\langle\Delta E\rangle_i,
\ee
where $i=\{x,y,z\}$, $dN/dE$ is the total number of states inside the block per
unit  energy, and $\langle\Delta E\rangle_i$ is the mean fluctuation
in energy levels caused by replacing periodic by antiperiodic boundary
conditions in the  direction of the current.
The value of $\langle\Delta E\rangle_i$ is determined by the sensitivity of
the energy
levels to the boundary conditions, which  in turn depends on
the phase memory. Since  the
decoherence of the wavefunction occurs within the boundaries of the hyperblock,
$\langle\Delta E\rangle_i$ should be isotropic.
Also,  $\langle\Delta E\rangle_i$
is determined by the time it takes for a particle to cross
the system in a given direction\cite{Thouless}.  By  definition,
it takes  an electron the same dephasing time $\tau_{\varphi}$  to cross
the hyperblock
in each direction, so that $\langle\Delta E\rangle_i\sim \hbar/\tau_{\varphi}$.

A macroscopic block $\{L_x,L_y,L_z\}$  obtained
by fitting together  $N^3$   hyperblocks
$(L_x/\ell_x =L_y/\ell_y=L_z/\ell_z=N\gg1)$  also
has isotropic conductance $G\approx Ng$ which can be expressed in terms
of the components of the conductivity tensor: $G=\sigma_xL_yL_z/L_x=
\sigma_yL_xL_z/L_y=\sigma_zL_yL_x/L_z$.
As a result, we arrive to the following relationship between
conductivities:
\be
\frac{\sigma_x}{\sigma_y}=\frac{\ell_x^2}{\ell_y^2};\;\;
\frac{\sigma_x}{\sigma_z}=\frac{\ell_x^2}{\ell_z^2}.
\ee
This is a general result, valid for all media, not only for layered crystals.
For example, in an anisotropic Fermi liquid  $\ell_i\propto
V_{F,i}\tau_{\varphi}$, where
$V_{F,i}$ is a component
of the Fermi velocity. Then, Eq. (2) reduces to the ratio of the effective
masses. In the case of  anisotropic diffusion, the Thouless lengths
$\ell_i^2\propto D_i\tau_{\varphi}$, where $D_i$ is a
component of the diffusion tensor.  For such a system, Eq. (2) is equivalent
to another known result $\sigma_i/\sigma_j=D_i/D_j$\cite{Altshuler,Wolfle}.
Finally, for
the case of variable range hopping (VRH),  
Eq.  (2) was obtained through a  different and less general approach, where
$\ell_i$ is the average hopping distance in the respective direction  \cite{Levin}.

We apply Eq. (2) to  highly anisotropic layered crystals in which electrons
lose coherence
in the c-direction over the smallest possible distance, the interlayer
spacing $\ell_0$
(unidirectional Ioffe-Regel limit, $\ell_z=const=\ell_0$).
For simplicity, the planes are considered to be isotropic; i.e.,
$\ell_x=\ell_y\equiv \ell$.
Under these conditions, Eq. (2) gives
\be
\sigma_c=\frac{\sigma_a\ell_0^2}{\ell^2}.
\ee
Thus, the out-of-plane conductivity, which has been one of the most
enigmatic features of the normal state of cuprates, is completely
determined by the
in-plane conductivity and  the in-plane  Thouless length.

Typically, quantum phenomena (e.g., superfluidity and superconductivity)
reveal themselves  macroscopically
through  phase coherence established over macroscopic distances.
Here,  anomalously  strong interlayer  {\it decoherence} in the normal
state (of still unknown origin)
has macroscopic consequences such as a strongly  T-dependent resistive
anisotropy which reflects the
T-dependence of the  in-plane  phase coherence length. Equation (3) cannot
be obtained from the quasiclassical
kinetic equation because c-axis transport is highly nonclassical even
when the in-plane transport  can be, in
principle, treated quasiclassically.

As a first  example of the applicability of Eq. (3) to layered crystals, we  show below that for
optimally and {\it nearly} optimally doped cuprates, Eq. (3) along with diffusive in-plane
transport gives a description  of the out-of-plane resistivity  which agrees very well with
experiment and explains the  observed correlation between  $\rho_a$ and $\rho_c$.  While  Eqs. 
(2) and (3)  are fundamental,   the T-dependence of the in-plane TL is not, since TL can
be determined by a variety of  dephasing processes with one of them
dominant in a certain T range  for
a given system.  Here we assume that the {\it strongest} dephasing process
is due to the thermal spread of
energies, so that TL is determined  by the thermal diffusion length.  [
Support for this assumption is provided
by Ref. \cite{Alt} which shows that the electron-electron interaction may
be equivalent to the  interaction of the
electrons with the thermal fluctuations of the electromagnetic waves.]
Then, the in-plane TL may be written as
\be
\ell^2=\xi^2+\frac{\hbar D}{k_BT}.
\ee
Here we  have added the  empirical cutoff length $\xi$ to the conventional
definition because  in a crystal, even
at high temperatures,  the phase coherence length cannot be shorter than a
certain finite length, e.g.  interatomic distance.   Since elastic  collisions do
not lead to the loss of coherence,  more
realistically
$\xi$ is comparable to the elastic mean free path [at least  far enough
from the metal-insulator  transition (MIT) as
we discuss later].

With the phenomenological expression for the in-plane resistivity,
$\rho_a=\beta_a+\alpha_aT$\cite{Cooper}, Eqs. (3) and (4) give:
\be
\rho_c=\beta_c +\alpha_c T+\frac{\gamma_c}{T},
\ee
where
\bea
\beta_c=\alpha_aT_0+\beta_a\frac{\xi^2}{\ell_0^2};\;\nonumber
\alpha_c=\alpha_a\frac{\xi^2}{\ell_0^2};\;\;
\gamma_c=\beta_aT_0,
\eea
and $T_0\equiv\hbar D/\ell_0^2k_B$. Two important aspects are worth noting.
First, a large constant $\beta_c$ appears even  in the absence of
the intercept $\beta_a$ (if $\rho_a=\alpha_aT$, then
$\rho_c=\beta_c+\alpha_cT$).  This apparent "residual resistivity" violates
 Matthiasen's rule,
because it does not scale to zero with the concentration of impurities.
Second, a finite positive intercept $\beta_a$ in
$\rho_a$ translates into a non-metallic term $\gamma_c/T$, so that
$\rho_c$ has a minimum.
Such correlations between $\rho_a$ and $\rho_c$ exist in
optimally and {\it slightly} underdoped  cuprates\cite{Cooper}.
From the data of
$YBa_{2}Cu_{3}O_{7-\delta}$\cite{Friedmann} for which
$\alpha_a\approx 0.5\;\mu\Omega\; cm/K$, $\;\beta_c\approx 1.1\;m\Omega\; cm$,
and $\alpha_c\approx 12.5\;\mu\Omega\; cm/K$, we estimate $T_0\approx 2200\;K$,
$D\approx 4\;cm^2/s$, and
$\xi=\ell_0(\alpha_c/\alpha_a)^{1/2}\approx 5\ell_0\approx 60\;\AA$
($\ell_0=11.7\;\AA$).
These are the least anisotropic crystals, in which a crossover from
incoherent to coherent
interlayer transport apparently takes place\cite{Basov}. Nevertheless,
their  anisotropy
is still strongly temperature dependent in agreement with Eqs. (3) and (4),
and changes by a factor of 2 - 3 between  room temperature and $T_c$. For
other types of
optimally doped layered cuprates,
which are  much more anisotropic,   the condition of  interlayer incoherence,
leading to Eq. (3),  is satisfied even better.

As a second  example of the applicability of Eq. (3) to cuprates, we consider
the other  doping extreme - insulating
crystals like $PrBa_{2}Cu_{3}O_{7-\delta}$\cite{Levin}. The incoherence
of the c-axis transport in this case means that the localized states are
two-dimensional (2D).  In the hopping regime, TL is  given by the average
hopping distance.
In the VRH regime
uncomplicated by Coulomb interactions, the average in-plane hopping
distance $R$
can be obtained by maximizing the hopping probability
\be
P(R)\propto \exp\left \{-2 \frac{R}{\lambda} -\frac{A}{{\cal
N}(R^2-R_0^2)T}\right \},
\ee
where  $\lambda$ is the localization length,
${\cal N}=const$ is the 2D density of states, and $A$ a numerical coefficient.
The only modification of the  traditional treatment of VRH is the
denominator $R^2-R_0^2$,
instead of $R^2$, which takes  into account that two localized states with
close energies cannot strongly overlap. If they do overlap,
the phonon interaction that causes hopping will also  hybridize them and
push apart
the energies of the new states. Strictly speaking, Eq. (6) is valid only
for $R\gg R_0\sim 2\lambda$. In this limit, the average hopping distance,
which in the hopping regime
determines  the in-plane TL, is $\bar R+ 2R_0^2/3\bar R$, where $\bar
R=(\lambda A/{\cal N} T)^{1/3}$.
Then, according to Eq. (3),
$\sigma_a/\sigma_c\approx [4R_0^2/3 +(\lambda A/{\cal N}T)^{2/3}]/\ell^2_0$.
This T-dependence of the anisotropy  ($a+bT^{-2/3}$)
was observed in insulating $PrBa_{2}Cu_{3}O_{7-\delta}$ and superconducting
$Y_{1-x}Pr_{x}Ba_{2}Cu_{3}O_{7-\delta}$\cite{Levin}.
Note that    with increasing localization length $\lambda$ (on approaching
the MIT),  the value of the anisotropy
increases while its temperature dependence weakens [$b/a\propto \lambda^{-4/3}$].

In the previous two examples we considered systems far away from the
metal-insulator transition.
The localization length on the insulating side and
the correlation 
\begin{figure}
\epsfxsize=\columnwidth \epsfbox{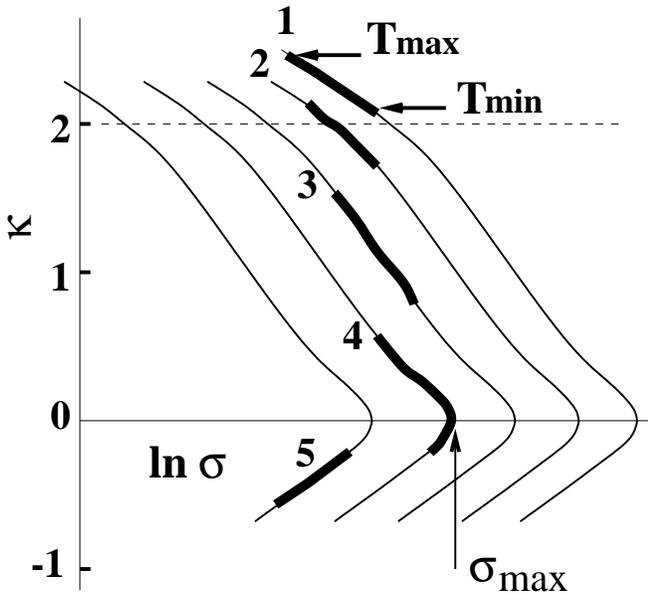}
\caption{Sketch of the trajectories $\kappa(\sigma_a )$ vs.  $\ln\sigma_a$.
Bold segments  correspond to  "experimentally accessible" 
values of conductivities as explained in the text. 
The thin lines are hypothetical extensions
of the trajectories drawn under the assumption that the segments
corresponding to different doping levels are parts of a continuous
curve $\kappa(\sigma/\bar\sigma)$, shifted with respect to each other
due to different normalization constants $\bar\sigma$. 
The curves as shown do not represent any particular set of measurements, but rather a
very simplified view of the overall behavior transpiring from available 
data.}
\label{alldata}
\end{figure}
\noindent
length  on the metallic
side  of the MIT  increase on approaching the  transition, and therefore,
at low temperatures, the phase coherence  length near MIT  is dominated by the
divergent localization/correlation length, respectively.  As a result,
according to Eq. (3), at a given T,
the anisotropy  $\rho_c/\rho_{ab}$ changes nonmonotonically with doping,
reaching a maximum in the  sample  closest to the MIT.
Such a nonmonotonic variation of $\rho_c/\rho_{ab}$ with doping $x$  is
observed  in $Y_{1-x}Pr_{x}Ba_{2}Cu_{3}O_{7-\delta}$ crystals\cite{C.N.Jiang}.
At the same time,  within a given range of temperature,  the T-dependence
of TL (and, hence,  anisotropy)
near MIT becomes substantially weaker than in either
metallic [$\rho_c/\rho_a\sim
1/T$ from Eqs. (3) and (4)]   or strongly insulating ($\rho_c/\rho_a\sim
T^{-2/3}$) phase.
This explains why  the crystals of Refs.\cite{Ando,Ando1}, which are
strongly underdoped and much closer to
the MIT than those of  Ref.\cite{Friedmann}, exhibit a  much weaker
T-dependence of resistivities and
anisotropy at low temperatures.

To extend the quantitative description of conductivity to  all  levels of
doping,
we use the following idea: the {\it isotropic} conductance $g$ [ Eq. (1)] of the
hyperblock with the sides equal to the respective TL $\{\ell,\ell,\ell_0\}$
can be described as a function of $\ell$ only, so that
the temperature and  magnetic field  dependence
of $g$ appears {\it exclusively} through that of $\ell$.
The justification of this hypothesis
can be partly found in the ideas of the scaling theory\cite{Lee,Abrahams}.
Indeed, it is  known\cite{Anderson} that the conductivity of a
conventional 3D Fermi liquid,
as well as the quantum corrections
\cite{Lee} can be expressed as
functions of TL with no explicit temperature dependence.
For a medium "composed" of hyperblocks $\{\ell,\ell,\ell_0\}$,
the in-plane conductivity $\sigma_a$ differs from conductance $g$
by a constant; i.e., $\sigma_a(\ell )=g(\ell )/\ell_0$. Therefore,
below we  discuss the $\sigma_a(\ell )$ dependence instead of $g(\ell )$.
This assumption that the T-dependence of the conductivities enters  only
through that of $\ell$, immediately explains  why
metallic $\sigma_a$ and nonmetallic $\sigma_c$ can coexist.
For example, if $\sigma_a\propto \ell^{\nu}$ with
$0<\nu < 2$, then, according to Eq. (3),
$\sigma_a$ is metallic ($d\sigma_a/d\ell >0$) while $\sigma_c$ is
nonmetallic ($d\sigma_c/d\ell <0$).

A quantitative description of  the $\sigma_a(\ell )$ dependence can be
given through
the  logarithmic derivative
\bea
\kappa=\frac{d\ln\sigma_a }{d\ln\ell }\nonumber
\eea
which we  treat  as a function of $\sigma_a$. This definition formally
resembles the "trajectories" of the scaling theory\cite{Abrahams}.
The difference is that  Ohm's law does not impose any restrictions
on the $\sigma_a(\ell)$ dependence  and  the limiting values of $\kappa$.
Instead, we  rely on  experimental data to infer
the shape of the trajectory $\kappa (\sigma_a )$. We can do this because,
according to
Eq.(3), the measured  anisotropy $\eta\equiv (\rho_c/\rho_a)^{1/2}$ gives
directly
the temperature dependence of TL; i.e., $\ell=\eta \ell_0$. Therefore,  the
experimental
dependence  $\sigma_a$ vs. $\eta$ determines $\sigma_a(\ell)$ and
$\kappa(\sigma_a)=d\ln\sigma_a/d\ln\eta $ dependences.
The thick segments in  Fig. 1  are  schematic representations of $\kappa(\sigma_a )$ for 
%
several levels of doping, where the range of $\sigma_a $ corresponds to
"accessible"
temperatures  $T_{min}<T<T_{max}$ ($T_{max}$ is typically $300-350\;K$
and $ T_{min}\approx T_c$ in superconducting crystals). The thin lines
indicate the
hypothetical extensions of the trajectories outside of this range.
According to Eq. (3),
\bea
\frac{d\ln\sigma_c }{d\ln\ell }=\kappa -2.   \nonumber
\eea
Therefore, both $\sigma_a$ and $\sigma_c$ are metallic for $\kappa>2$ and
nonmetallic for $\kappa<0$, while metallic
$\sigma_a$ and  nonmetallic  $\sigma_c$ coexist for $0<\kappa<2$.

Segment 1 in Fig. 1 represents optimally or overdoped regimes where both
conductivities $\sigma_a$ and $\sigma_c$ are metallic
at all temperatures $T>T_{min}$
because the whole segment is located above the threshold $\kappa =2$.
The previous
example of  optimally doped cuprates [$\rho_a=\alpha_aT$,  $\ell$  given by
Eq. (4), and
$\rho_c=\beta_c +\alpha_cT$] leads to $\sigma_a=q(\ell^2-\xi^2)$ and
$\kappa(\sigma )= 2+\bar\sigma/\sigma$, where $q=k_B/\alpha_a \hbar D$ and
$\bar\sigma=2q\xi^2$.

Segment 2 corresponds to a slightly underdoped system.
At high temperatures (small $\sigma_a$) $\kappa >2$ and,  therefore,
$\sigma_c$ is metallic. With  increasing $\ell$ (decreasing $T$) $\sigma_c$
increases,   reaches a
maximum  when $\kappa (\sigma_a )=2$, and then decreases.
Therefore, $\rho_c$  changes from metallic at high T to nonmetallic at low
T,  similar to  the  T-dependence  given by Eq. (5).
Within the same range of  temperature,  $\rho_a$ is metallic.

Segment 3 represents a moderately underdoped system.
It lies entirely within the range $0<\kappa
<2$ and  corresponds to metallic $\rho_a$ and nonmetallic $\rho_c$
for all   $T_{min}<T<T_{max}$.

Segment 4 corresponds to strongly underdoped
crystals\cite{Levin,C.N.Jiang} with $\sigma_a$ changing from
metallic at
high T ($\kappa >0$) to nonmetallic at lower T ($\kappa <0$).
The singularity $\kappa =0$ is integrable:
\be
\kappa (\sigma )
\approx\pm \zeta\left (\ln\frac{\sigma_{max}}{\sigma}\right )^{1/2},
\ee
so that $\sigma_a(\ell)$ as determined by the equation
\be
\int_{\sigma_{max}}^{\sigma_a}\frac{d\ln\sigma}{\kappa (\sigma
)}=\ln\frac{\ell}{\ell_1}
\ee
reaches the maximum value $\sigma_{max}$ at a finite $\ell=\ell_1$ and
decreases with
further increasing $\ell$ (decreasing $T$).
Equations (7) and (8) give
$\sigma_a(\ell)\approx \sigma_{max}\exp\{-\zeta^2\ln^2(\ell/\ell_1 )/4\}$.

Finally, segment 5 corresponds to an insulating crystal.
In this case, $\ell\approx\bar R$ and  $\sigma_a\sim
\sigma_0\exp\{-3\ell
/\lambda\}$. Therefore,  $\kappa(\sigma )\approx \ln(\sigma/\sigma_0)$.

As shown, all five  curves in Fig. 1 represent the same $\kappa
(\sigma/\bar\sigma )$ dependence,
shifted with respect to each other due
to different values of $\bar\sigma $  (determined
by the  density of carriers $n$ and
other  parameters which cannot be absorbed into TL). The reduction of $n$
by doping reduces $\bar\sigma $, shifting the respective
segment of $\kappa (\sigma/\bar\sigma )$ to lower absolute values of
$\sigma$ and, at the same time, ahead along the trajectory in terms of the
reduced variable $\sigma/\bar\sigma$.
Integration similar to Eq. (8) gives  $\sigma(\ell )$ of the form
\bea
\frac{\sigma_a}{\bar\sigma }=f\left (\frac{\ell}{\bar\ell}\right ),
\eea
where $f(y)$ is a universal function [$f(1)=1$], so that
the conductivity of the crystals at different levels of doping
is described by the same $f(\ell)$ dependence, provided the appropriate
choice of doping dependent parameters $\bar\sigma$ and $\bar\ell$.
This hypothesis can be verified experimentally by
checking whether the conductivity
at different doping levels
plotted against anisotropy $\eta$ form a continuous curve when $\sigma_a$
and $\eta$
are properly normalized.

This idea of universality of the $\sigma_a (\ell )$ dependence
allows the prediction of the behavior  of the normal state
conductivities at temperatures below  $T_c$.
For example,  let segment 3 in Fig. 1 represent   a
superconducting crystal with  metallic $\sigma_a$ and nonmetallic
$\sigma_c$ for all $T_c<T<T_{max}$.
If we suppress  the onset of superconductivity with a magnetic field  in order
to reveal the normal state  at $T<T_c$, we will observe that
the trajectory crosses the  threshold  $\kappa =0$ and $\sigma_a$ also
becomes nonmetallic.
This type of development has been observed \cite{Ando,Ando1} in
underdoped cuprates.

It is possible that the trajectories  shown in Fig. 1 oversimplify the
situation, and there may be more than one type of $f(\ell)$
dependence. The data suggest that in $Y_{1-x}Pr_{x}Ba_{2}Cu_{3}O_{7-\delta}$, a
metal-insulator transition takes place at
intermediate  doping levels  somewhere between optimally doped and strongly
insulating phases\cite{C.N.Jiang}.  As a result,  there may be at least two
types
of trajectories (or a branching point), corresponding to
the phases on either side of the MIT.

The assumption  that the response of the conductivities
to external perturbations other than temperature also comes from that of TL
allows  us to predict these responses because
\be
\frac{\partial\ln\sigma_a}{\partial x}=\kappa \frac{\partial\ln
\ell}{\partial x};\;\;
\frac{\partial\ln\sigma_c}{\partial x}=(\kappa -2)\frac{\partial\ln
\ell}{\partial x}.
\ee
Here $x$ can be magnetic field or concentration of impurities.
For $\kappa >2$ or $\kappa < 0$ both conductivities change similarly.
However, for a wide range of doping levels and temperatures [Fig. 1],
for which $0<\kappa<2$, the two conductivities respond oppositely.

For example,  in $Bi_2Sr_2Ca(Cu_{1-x}Zn_x)_2O_{8+y}$ single crystals $\sigma_c$
{\it increases}   and $\sigma_a$ decreases with increasing  $Zn$
concentration\cite{Jeon}.
The impurities
reduce the in-plane TL ($\partial\ell/\partial x <0$) by decreasing the
elastic mean free path and
the diffusion coefficient.  The crystals used in\cite{Jeon} exhibit metallic
$\rho_a(T)$ and nonmetallic $\rho_c(T)$, so that $0<\kappa <2$
and, as a result,  $\partial\sigma_a/\partial x <0$ while
$\partial\sigma_c/\partial x >0$.
Sometimes, the increase of the c-axis conductivity in response to a perturbation
has been interpreted in  literature as a crossover to coherent transport in the
c-direction. We see that this is not necessarily the case. The
anisotropy may decrease and $\sigma_c$ increase due to the reduction
of the in-plane TL, even when the c-axis TL remains fixed.

The effect of the magnetic field $H$ on conductivities
is determined by the  destructive interference\cite{Lee},
resulting in the reduction of TL ($\partial \ell/\partial H <0$).
Therefore,  according to
Eq. (3), the {\it magnetoanisotropy} [$\delta (\rho_c/\rho_a )$]
should be negative, while the sign and relative magnitude of the
magnetoconductivities $\Delta\sigma_a$ and $\Delta\sigma_c$
depend on the value of $\kappa$ according to Eq. (10).
One can see that the magnetoresistivities correlate with the temperature
coefficient of
the respective component of the resistivity:
\bea
\frac{\partial\rho_{a,c}}{\partial H}=Q \frac{\partial\rho_{a,c}}{\partial
T},\nonumber
\eea
where $Q=(\partial\ell /\partial H )/(\partial\ell /\partial T)$. Since $Q>0$,
the sign of magnetoresistance
$\Delta\rho_{a,c}$ is the same as
the sign of $\partial\rho_{a,c}/\partial T$,  as indeed was reported in
Refs.\cite{Ando,Ando1,Yan}.
The magnitude of the magnetoeffects
depends on the number of flux quanta ($h_{i}$)
permeating  the hyperblock. For $H\| c$,  $h_c=
H\ell^2/\phi_0= (\rho_c/\rho_a)H/H_0$
($H_0\equiv \phi_0/\ell_0^2 $ and $\phi_0$ is the quantum of flux).
If the field is parallel to the planes, $h_a=H\ell\ell_0/\phi_0=
(\rho_c/\rho_a)^{1/2}H/H_0$. Thus, $h_c\gg h_a$ which may result in a
strong angular dependence of the magnetoeffects.

{\it This research was supported  by the National Science Foundation under
Grant No.  DMR-9801990}.

\end{document}